\documentclass[aps,prl,twocolumn,floatfix,10pt,amssymb,amsfont,amsmath,superscriptaddress,float,longbibliography]{revtex4-2}
\usepackage{graphicx}
\usepackage{dcolumn}
\usepackage{bm}
\usepackage{hyperref}
\usepackage{graphicx,amsmath, amssymb} 
\usepackage{graphicx} 
\usepackage{amsmath}
\usepackage{braket}
\usepackage{xcolor}
\usepackage{soul}\setstcolor{red}
\usepackage{comment}
\usepackage{dsfont}
\usepackage{layouts,braket}

\begin{document}
\title{Global Tensor Network Renormalization for 2D Quantum systems:\\ A new window to probe universal data from thermal transitions}

\author{Atsushi Ueda*}
\author{Sander De Meyer*}
\author{Adwait Naravane}
\author{Victor Vanthilt}
\affiliation{Department of Physics and Astronomy, Ghent University, Krijgslaan 281, 9000 Gent, Belgium}

\author{Frank Verstraete}
\affiliation{Department of Physics and Astronomy, Ghent University, Krijgslaan 281, 9000 Gent, Belgium}
\affiliation{Department of Applied Mathematics and Theoretical Physics, University of Cambridge,\\ Wilberforce Road, Cambridge, CB3 0WA, United Kingdom}

\date{\today}

\begin{abstract}
We propose a new tensor network renormalization group (TNR) scheme based on global optimization and introduce a new method for constructing the finite-temperature density matrix of two-dimensional quantum systems. Combining these two into a new algorithm called thermal tensor network renormalization (TTNR), we obtain highly accurate conformal field theory (CFT) data at thermal transition points. This provides a new and efficient route for numerically identifying phase transitions, offering an alternative to the conventional analysis via critical exponents.
\end{abstract}

\maketitle

\paragraph{Introduction} 
Understanding emergent phenomena in quantum many-body systems is a central goal of condensed matter physics \cite{PhysicsPhysiqueFizika.2.263_kadanofforiginal, WilsonNRG}. These phenomena typically involve strong correlations that lie beyond the reach of conventional mean-field or perturbative methods. To investigate such systems, researchers often rely on lattice models. However, due to the exponential growth of the Hilbert space with system size, exact diagonalization is limited to small systems.  To overcome this challenge, a variety of numerical techniques have been developed. One of the earliest and most influential is the numerical renormalization group (NRG)~\cite{WilsonNRG}, introduced by Wilson. NRG is based on a recursive blocking of local Hamiltonians combined with truncation of the Hilbert space, enabling simulations beyond the reach of exact diagonalization. It marked a significant milestone by successfully solving the Kondo problem and demonstrating the power of tensor-network-based approaches. Nevertheless, NRG's accuracy remained limited, particularly for systems exhibiting extended entanglement. This shortcoming was later addressed by the density matrix renormalization group (DMRG)~\cite{Whitedmrg}, introduced by White, which dramatically improved both accuracy and efficiency for one-dimensional quantum systems. The key innovation of DMRG is that it incorporates the Hamiltonian of the entire system when performing truncation using so-called ``environments", which encode the rest of the (infinite) system in a finite manner. This perspective leads to a notion of {\it global} optimization using the environment and inspired several one-dimensional variational algorithms~\cite{TEBD,iTEBD,PBCdmrg,idmrg,PhysRevB.97.045145_vumps}. 

In recent years, significant efforts have been devoted to developing projected entangled pair state (PEPS) algorithms for two-dimensional quantum systems \cite{Cirac_2009, Verstraete_cirac_2008, PhysRevB.83.245134_cirac}. In stark contrast to the one-dimensional case, these systems pose substantial challenges. For example, even computing the norm of a non-uniform random PEPS becomes computationally intractable, as the exact contraction is a \#P-hard problem \cite{schuchComputationalComplexityProjected2007}. This calls for the development of efficient approximation schemes.
Tensor renormalization group (TRG)~\cite{levinTensorRenormalizationGroup2007,xieCoarsegrainingRenormalizationHigherorder2012} and tensor network renormalization (TNR)\cite{guTensorEntanglementFilteringRenormalizationApproach2009,evenblyTensorNetworkRenormalization2015,  evenblyAlgorithmsTensorNetwork2017,yangLoopOptimizationTensor2017, Bal_tnrplus, hauruRenormalizationTensorNetworks2018,Homma_nuclear, LoopTNRFermions} have emerged as a key framework for addressing this challenge and have seen considerable progress over the past decade. However, most existing TNR algorithms rely on local optimization procedures, which inherently limit the accuracy of simulations in two dimensions. Overcoming this limitation remains a critical objective for advancing the simulation of two-dimensional quantum systems. 

In this Letter, we address this issue by proposing a new TNR algorithm based on global optimization. We demonstrate that our global contraction scheme significantly improves accuracy compared to conventional methods. Building on this foundation, we introduce an efficient contraction method for three-dimensional tensor networks representing finite-temperature density matrices of a two-dimensional quantum system. Our approach enables the extraction of conformal field theory (CFT) data from thermal transitions. Notably, previous attempts to simulate two-dimensional quantum systems under periodic boundary conditions (PBC)~\cite{Gleb2025,Shaojun2025} or at finite temperature \cite{Czarnik_iterative, XTRG, FCLS_Corboz, tangent_trg_2023} have faced significant obstacles. Our method overcomes both challenges simultaneously, opening a new avenue for exploring quantum critical phenomena in two dimensions.

\paragraph{TNR with global optimization} 

\begin{figure}[tb]
    \centering
    \includegraphics[width=86mm]{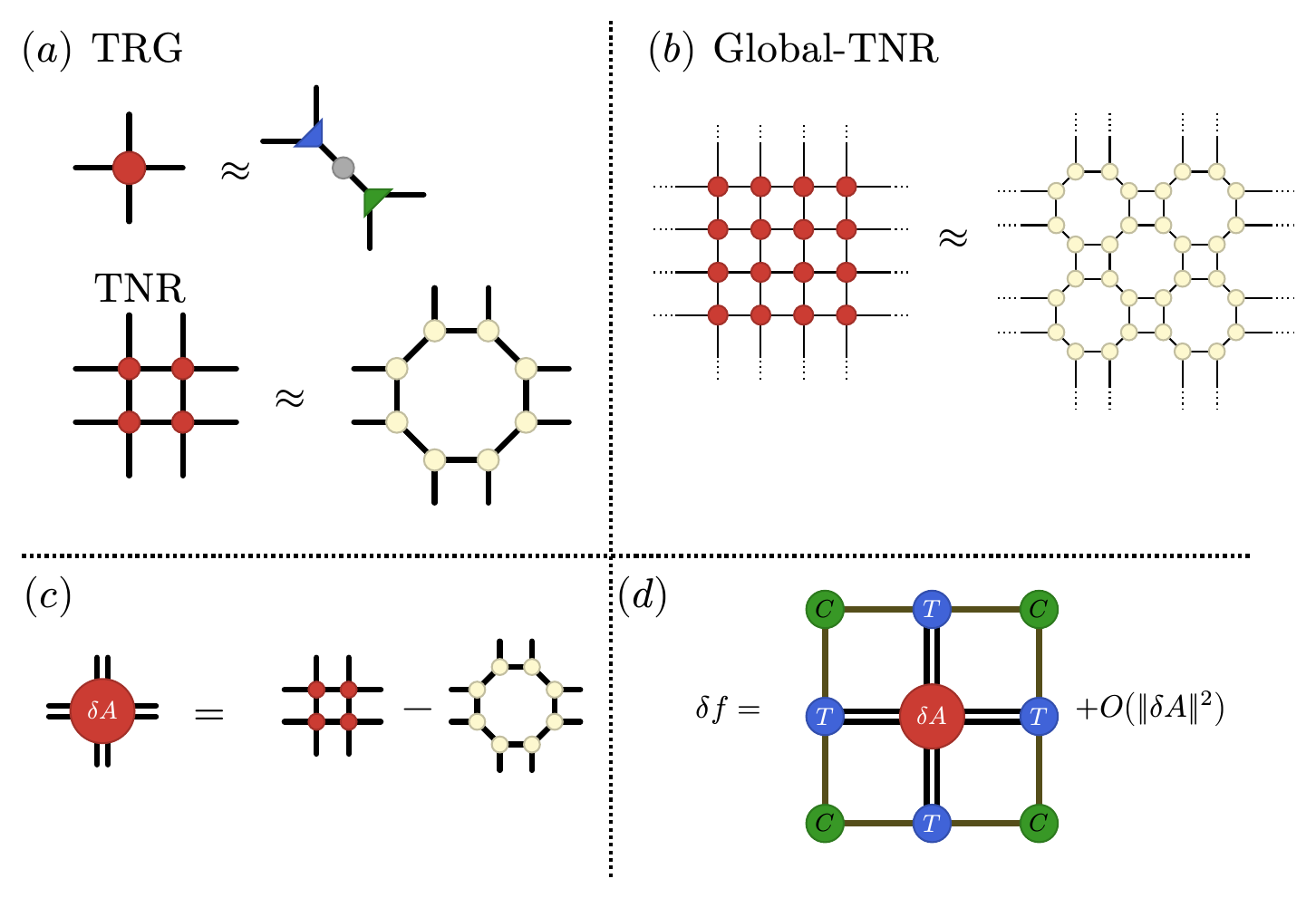}
    \caption{$(a)$ The objective of optimization during tensor decompositions for TRG and TNR. TNR improves the accuracy by incorporating a larger unit cell. $(b)$ Global TNR decomposes tensors to minimize the difference of the whole tensor networks. $(c)$ A pictorial definition of $\delta A$. It is defined as a difference of the eight-leg tensor. $(d)$ The difference of the whole tensor network $\delta f$ is approximated by the expectation value of the error with respect to the environment computed from CTMRG to first order.}
    \label{fig:illustration}
\end{figure}

TRG, developed by Levin and Nave~\cite{levinTensorRenormalizationGroup2007}, is a widely used method for computing the partition function of two-dimensional classical statistical systems. In TRG, the partition function is expressed as a contraction of four-leg tensors encoding the local Boltzmann weights. Since the exact contraction of a two-dimensional tensor network is exponentially costly, TRG performs successive coarse-graining steps that reduce the number of tensors by half at each iteration through a sequence of decompositions and recombinations. After several such steps, the network consists of only a few tensors, which can then be contracted exactly. To keep the procedure numerically tractable, TRG applies truncation during each decomposition step using singular value decomposition (SVD), minimizing the error in the local approximation. This procedure is illustrated in the top-left panel of Fig.~\ref{fig:illustration}. 

TNR can be viewed as a generalization of the local approximation in TRG to a two-by-two unit cell. This extension yields more accurate approximations for larger system sizes and improves the overall precision of the tensor network contraction as it incorporates the crucial feature of \textit{entanglement filtering}. When using two-by-two unit cells, redundant loop-like degrees of freedom can appear on the plaquettes formed by the four tensors. These structures are known as \textit{corner double line} (CDL) tensors. Gu and Wen proposed a way to remove these degrees of freedom called entanglement filtering \cite{guTensorEntanglementFilteringRenormalizationApproach2009}. Other prominent TNR algorithms, like Loop TNR, also include an entanglement filtering step \cite{yangLoopOptimizationTensor2017}. TRG fails to remove these non-universal, short-range contributions . From the renormalization group perspective, such CDL structures encode short-scale physics and should be systematically removed during the coarse-graining process. This is precisely what TNR achieves. 

Despite their advantages, most TNR schemes still rely on \textit{local} optimization. Just as DMRG improves upon NRG by incorporating the global environment, TNR can be further enhanced by considering the full tensor network during decomposition. The key idea is to minimize the {\it{global}} error introduced when decomposing a tensor, as illustrated in the top-right panel of Fig.~\ref{fig:illustration}. In this setup, we extend beyond a two-by-two unit cell, with the surrounding tensors indicated by dotted lines.

Unfortunately, evaluating the full contraction difference exactly is intractable. Instead, we perform a Taylor expansion of the global error with respect to the variation in the two-by-two unit cell, denoted by $\delta A$. In the infinite system limit, the first-order term in $|\delta A|$ corresponds to the expectation value (i.e., the one-point function) of $\delta A$, which can be efficiently approximated using environment tensors.

We compute these environment tensors using the corner transfer matrix renormalization group (CTMRG)~\cite{nishino_kouichi_okunishi_1996_ctmrg}, applied to a two-site unit cell. We denote the resulting environment tensor as $\Gamma_{\mathrm{env}}$. To ensure the validity of the Taylor expansion and prevent uncontrolled deformations, we also include the Hilbert-Schmidt norm of $\delta A$ in the cost function.

The global optimization amounts to minimizing the following cost function:
\begin{align}
C(A) = \|\delta A\|_F^2 + \alpha \|\Gamma_{\mathrm{env}} \delta A\|_F^2,
\end{align}
where $\alpha$ is a weighting parameter controlling the contribution of the environment term and $\| \cdots\|_F$ is the Frobenius norm. In this study, we set $\alpha = 1$.

During the optimization, we impose spatial $C_4$ rotational symmetry and reflection symmetries along the $x$ and $y$ axes within the two-by-two unit cell. These symmetries relate the white three-leg tensors of Fig. \ref{fig:illustration} and effectively fix their gauge. This gauge fixing improves numerical stability, particularly in critical systems~\cite{yangLoopOptimizationTensor2017,Liclock,ueda2023fixedpoint}. After entanglement filtering, the cost function can be efficiently minimized using standard gradient-based optimization techniques.
\begin{table}[t]
    \centering
    \caption{The free energy of the classical 2d Ising model at $T=T_c$ compared with the exact solution.}
    \begin{ruledtabular}
    \begin{tabular}{c|c|c|c|c|c}
      $\delta f$&TRG &GTRG&CTMRG&Loop TNR  & Current work \\
      \hline
         Ising&$9.8\cdot 10^{-6}$& $6.9\cdot 10^{-7}$& $2.7\cdot 10^{-7}$& $3.5\cdot 10^{-8}$& $8.5\cdot 10^{-10}$
    \end{tabular}
    \end{ruledtabular}
    \label{tab:f_ising}
\end{table}
\begin{figure}[tb]
    \centering
    \includegraphics[width=86mm]{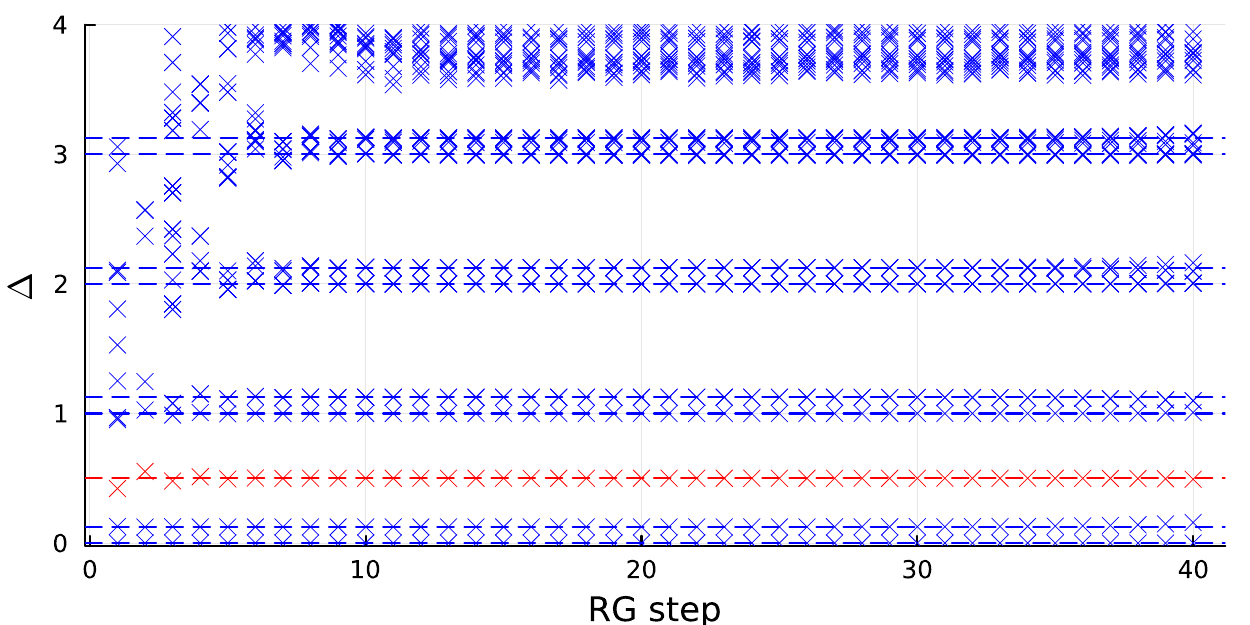}
    \caption{The CFT spectrum obtained from the transfer matrix spectrum of the classical Ising model at critical temperature. The red and blue crosses correspond, respectively, to the central charge and the scaling dimensions.}
    \label{fig:cft_2dising}
\end{figure}
Table~\ref{tab:f_ising} presents our benchmark results for the classical Ising model at criticality using bond dimension $\chi = 16$. We achieve an improvement of one to two digits in the accuracy of the free energy. Moreover, as shown in Fig.~\ref{fig:cft_2dising}, the critical spectrum remains stable up to 40 renormalization steps. While similar ideas of global optimization have previously been applied to TRG~\cite{xiesrg,Morita_globalTRG}, they did not yield a stable critical spectrum. Our results demonstrate that the combination of global optimization with TNR is crucial for accurately capturing physical properties at criticality.

\paragraph{CFT data from thermal transitions} 
A key application of our method is the efficient simulation of two-dimensional quantum states at finite temperature. We represent the thermal density matrix \( \hat{\rho} = e^{-\beta H} \) as a contraction of a three-dimensional tensor network, where the Hamiltonian in the exponent is expressed as a projected entangled-pair operator (PEPO). Although the PEPO has a finite bond dimension, its exponential generally does not. However, for small values of $\beta$, the cluster expansion method yields an accurate PEPO representation \cite{vanheckeSymmetricClusterExpansions2021, vanheckeSimulatingThermalDensity2023a, demeyer2026}. We call the tensor that generates this translationally invariant PEPO an elementary tensor. By decomposing the imaginary-time evolution into small steps \( e^{-\frac{\beta}{N} H} \), the full thermal density matrix can be expressed as an \( N \)-fold product $\hat{\rho} = \left(e^{-\frac{\beta}{N} H}\right)^N$.
Consequently, evaluating the density matrix on a system of size \( (L_x, L_y, \beta) \) reduces to contracting a three-dimensional network of size \( (L_x, L_y, N) \) composed of elementary tensors.

The contraction of such three-dimensional networks is generally intractable, typically requiring computational costs on the order of \( O(\chi^7 \text{--} \chi^{11}) \). Alternatives include local contractions using HOTRG \cite{xieCoarsegrainingRenormalizationHigherorder2012} or ATRG \cite{adachiAnisotropicTensorRenormalization2020} and finding the boundary PEPS of the 3D tensor network \cite{vanderstraeten2018residual}. In contrast, our method achieves a significantly reduced cost of \( O(\chi^6) \), comparable to the complexity of HOTRG in a classical two-dimensional system.

In the following, we propose an efficient contraction scheme. First, we construct the elementary PEPO tensor using a cluster expansion for small $\Delta\beta$. Then, we stack these tensors sequentially along the imaginary-time direction to form a column tensor. At each step, an elementary tensor is contracted \textit{linearly} on top of the previously renormalized column tensor, followed by a truncation of the bond dimensions of the four bonds in the spatial direction via a projection~\cite{yangLoopOptimizationTensor2017,Wang2011ClusterUF,PhysRevB.100.035449}. This projection minimizes the cost function illustrated below:
\begin{align}
    \includegraphics[width=70mm]{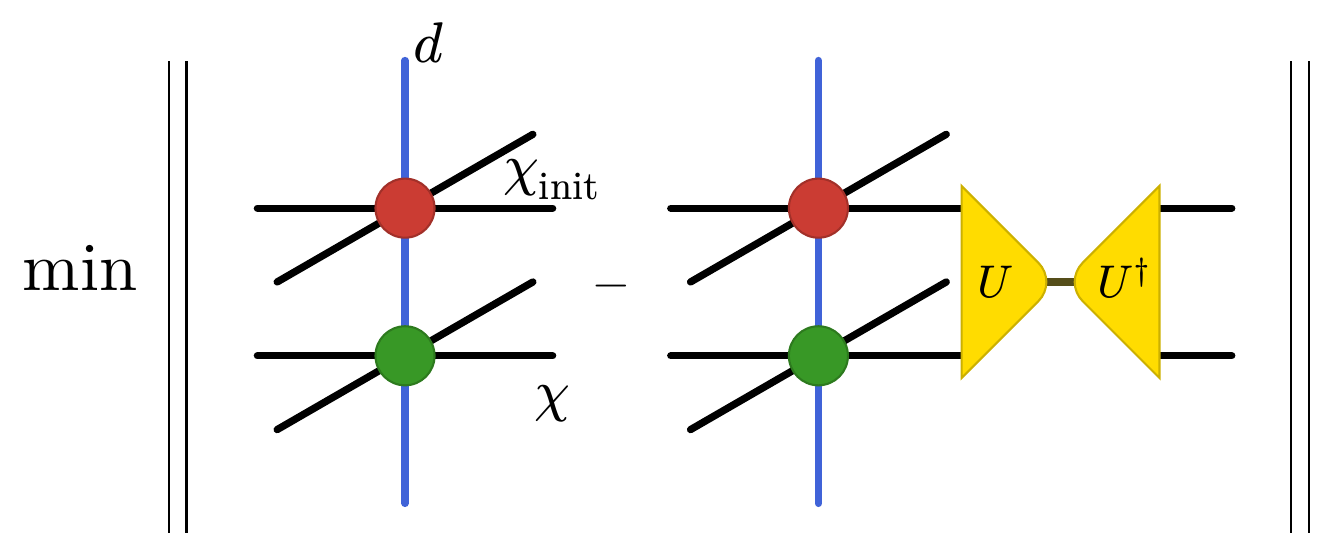} \; .
\end{align}
It is worth noting that the series of projections applied during the linear stacking results in a tensor network structure that resembles an MPS \cite{perez-garciaMatrixProductState2007a} along the imaginary-time axis~\cite{vtrg,vtrg2}(see also Fig. 8 of Ref.~\cite{vtrg2}). Further optimization of these projectors by DMRG-like sweeping can increase the accuracy \cite{VTNR_Czarnik}. After $N$ steps, we get a column PEPO representing $\beta = N \Delta\beta$. Second, we trace out the physical indices. The column tensor ends up becoming a 2D classical tensor. Finally, we perform TNR for $n$ steps in the $x$-$y$ plane. The partition function with PBC is obtained by contracting the open legs in the $x$ and $y$ directions. Moreover, the CFT information is accessible through the transfer matrix of the renormalized tensor.

We apply this scheme to the transverse field Ising model in two dimensions, defined as
\begin{align}
    H = -\sum_{\langle i, j\rangle}\sigma^z_i\sigma^z_j-\lambda\sum_i\sigma_i^x.
\end{align}

\begin{figure}[tb]
    \centering
    \includegraphics[width=86mm]{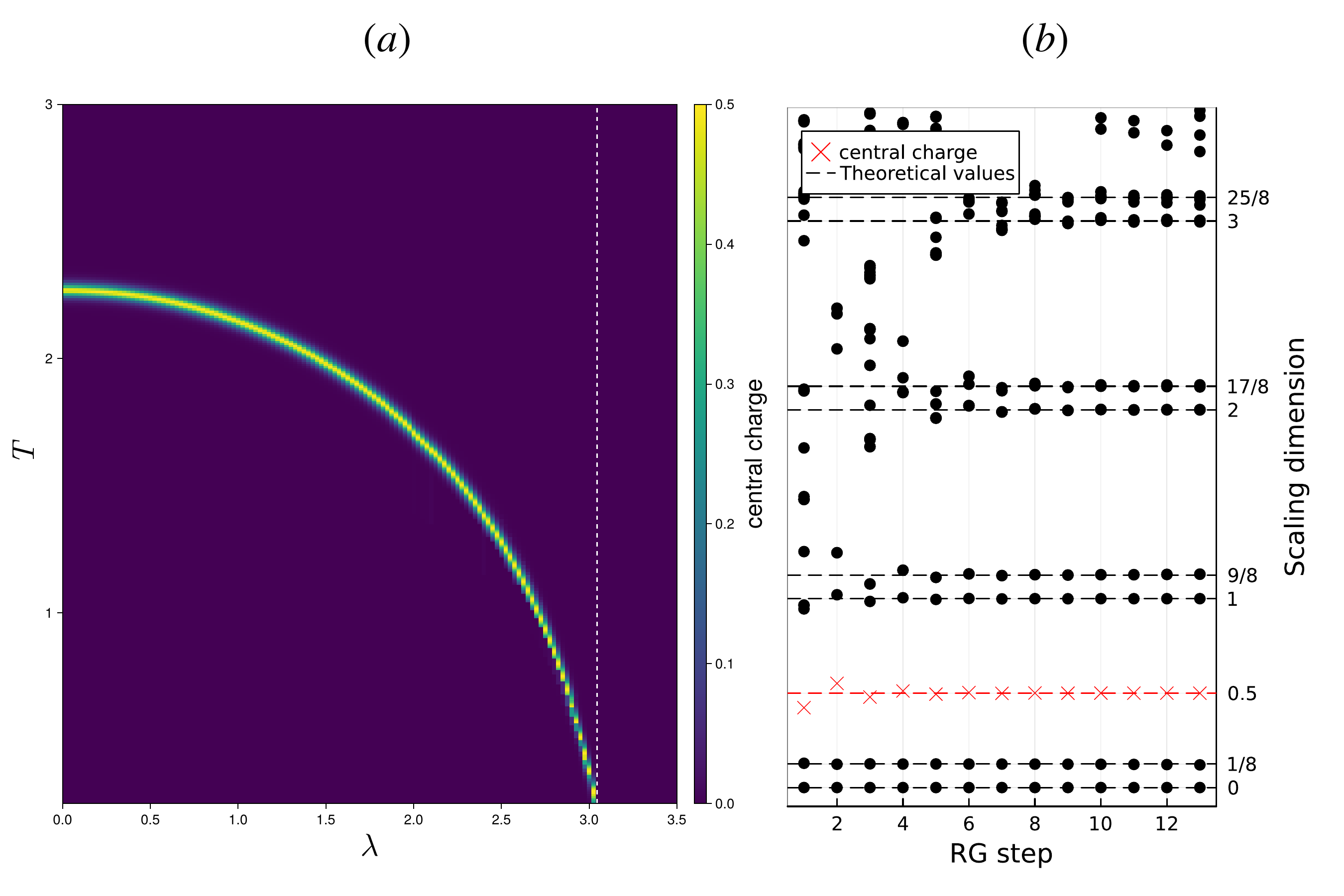}
    \includegraphics[width=86mm]{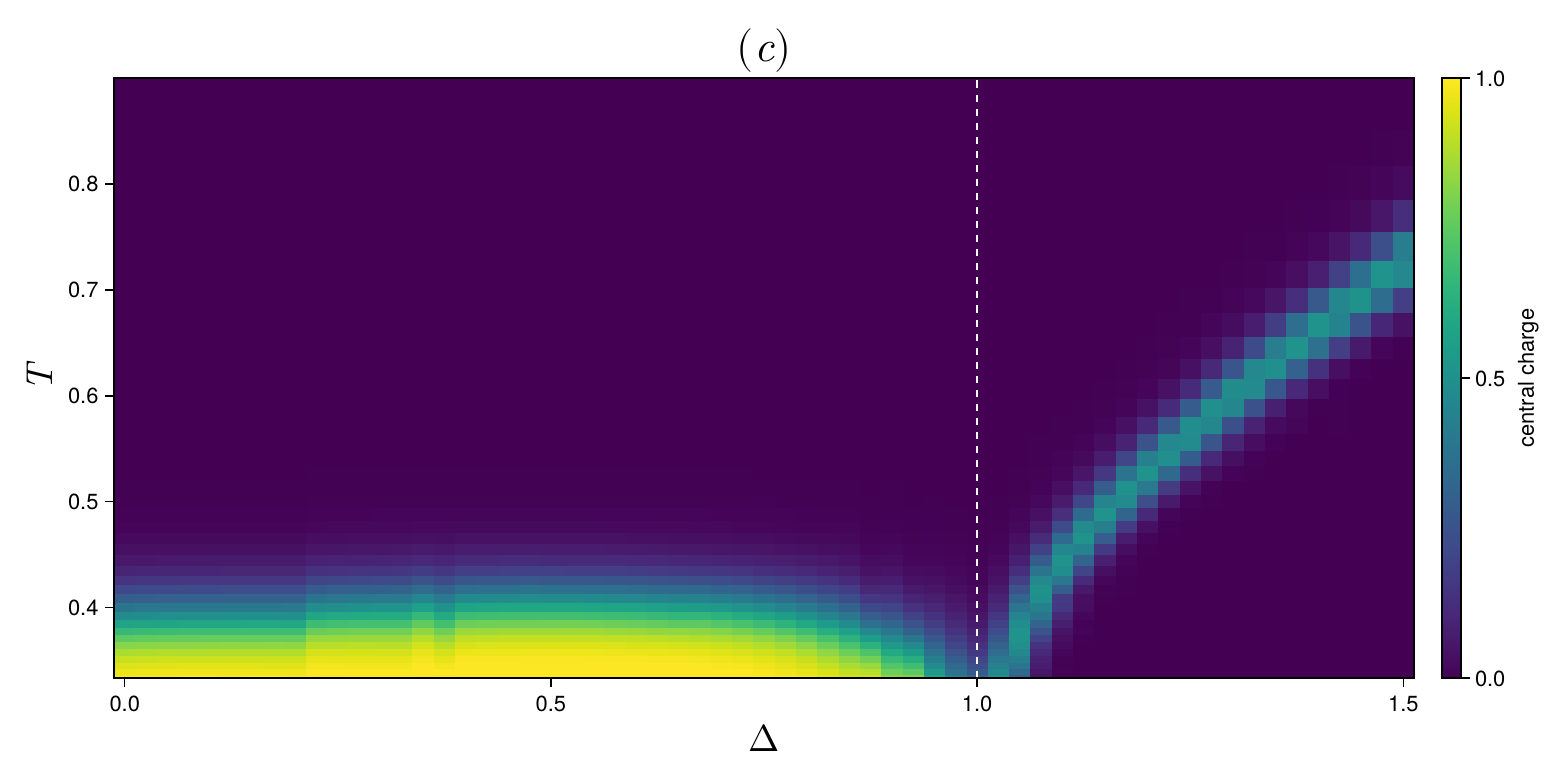}
    \caption{The finite-$T$ phase diagram of $(a)$ the transverse-field Ising model and $(c)$ the XXZ model on a square lattice. $(a)$ The result with $\chi_\tau = 12$ in the imaginary-time evolution and $\chi = 10$ in performing global TNR.The color map indicates the central charge, where $c=1/2$ at the thermal transition line. The transition line approaches the quantum critical point, indicated by a white dotted line, as $T$ approaches zero. $(b)$ The transfer matrix spectrum of the finite-$T$ density matrix at the transition point $(\beta,\lambda) = (0.472,1.1)$. The dotted lines indicate the theoretical Ising CFT spectrum. $(c)$ $\Delta<1$ and $\Delta>1$ respectively have the BKT and Ising transitions.}
    \label{fig:heatmap}
\end{figure}

This model exhibits a quantum phase transition at zero temperature, located at $\lambda_c \simeq 3.04438$ \cite{PhysRevB.57.8494_dmrgtemp, PhysRevE.66.066110_montecarlotemp}. At finite temperature, thermal phase transitions emerge and are characterized by the two-dimensional Ising CFT. Figure~\ref{fig:heatmap}(a) presents the finite-temperature phase diagram obtained from our method, where the color scale indicates the effective central charge extracted from the transfer matrix after 12 RG steps. Along the critical curve, the central charge is nearly exactly $c = 0.5$, consistent with the CFT prediction. Moreover, this 2D CFT curve approaches the quantum critical point in the $T\rightarrow0$ limit. 

In the right panel (b), we show the detailed spectrum of a transition point at $(\lambda, \beta) = (0.472, 1.1)$. The extracted scaling dimensions and central charge are highly accurate even at higher energy levels. For example, we obtain $ c = 0.49996$ at 12 RG steps. This result highlights the remarkable accuracy of our TNR-based approach for finite-temperature quantum systems.

Our method achieves this efficiency by keeping the bond dimension in the imaginary-time direction fixed at $\chi_\tau$ during the linear construction of the column tensor. This significantly reduces the computational cost compared to conventional three-dimensional TRG schemes \cite{xieCoarsegrainingRenormalizationHigherorder2012}.

Finally, the bottom panel (c) shows the finite-temperature phase diagram of the XXZ model with $\chi_\tau = 16$:
\begin{align*}
    H = \sum_{\langle i, j\rangle}(\sigma^x_i\sigma^x_j+\sigma^y_i\sigma^y_j+\Delta\sigma^z_i\sigma^z_j).
\end{align*}
The spatial contraction is performed using Loop-TNR~\cite{yangLoopOptimizationTensor2017} with $\chi = 28$. We use Loop-TNR here to demonstrate that the thermal TNR scheme is generic, and is not restricted to Global TNR. For $\Delta < 1$, the system enters a Tomonaga--Luttinger phase through a BKT transition, whereas for $\Delta > 1$ it undergoes a single Ising transition. The extracted central charges, $c = 1$ and $c = 1/2$, are consistent with these two universality classes. The transition temperature $T_c \simeq 0.34$ at $\Delta = 0$ is also consistent with previous results~\cite{xxz1,xxz2,xxz3}. This provides further evidence that thermal TNR is an efficient method for detecting the universal data of thermal phase transitions.
\paragraph{Conclusion and Outlook}
In this Letter, we develop a new TNR algorithm based on global optimization. Analogous to how DMRG improves upon NRG, our scheme achieves significantly more accurate contractions of two-dimensional tensor networks. This enhanced accuracy extends naturally to the simulation of two-dimensional quantum systems at finite temperature, represented as three-dimensional tensor networks.
Upon coarse-graining along the imaginary-time direction and tracing out physical indices, the three-dimensional network reduces to an effective two-dimensional network. Applying our TNR method to this tensor network enables direct extraction of CFT data from the transfer matrix. While it is known that entanglement filtering is essential for accuracy in TNR, this step is often neglected in variational approaches such as the gradient-based optimization of PEPS, which target the thermodynamic limit. In contrast, our method operates at finite system sizes, where universal data can be accessed with controlled errors with minor finite bond dimension effects~\cite{PhysRevB.89.075116,AtsushiTNR}.

Notably, our approach does not require full three-dimensional entanglement filtering as in 3D TNR, but two-dimensional filtering suffices~\cite{Lyu2024, Lyu2025}. This opens a practical route to studying thermal transitions in challenging models whose finite-temperature phase diagrams remain unresolved, such as the $J_1$–$J_2$ model, which we plan to investigate in future work.

Our current method focuses on two-dimensional criticality and spatial scale invariance, thereby naturally filtering out scale-invariant behavior from quantum critical points. In future developments, we aim to extend the method to capture genuine three-dimensional criticality, characterized by scale invariance in both spatial and imaginary-time directions.

\paragraph{Acknowledgement}
The authors thank Kevin Vervoort, Masahiko Yamada, Nick Bultinck, and Zheng-Cheng Gu for insightful discussions. 
A.U. thanks Yue Zhengyuan for the discussion on the XXZ model. A.U. was supported by BOF-GOA (Grant No. BOF23/GOA/021). S.D.M. was supported by the Research Foundation Flanders (FWO) (doctoral fellowship No. 11A3C25N). V.V. was supported by the FWO (doctoral fellowship No. 1196525N). This work was supported by EOS (grant No. 40007526), IBOF (grant No. IBOF23/064), and BOF- GOA (grant No. BOF23/GOA/021). Computational resources (Stevin Supercomputer Infrastructure) and services used in this work were provided by VSC (Flemish Supercomputer Center), funded by Ghent University, FWO, and the Flemish Government.

\paragraph{Data avaliability}
 The source code will be available on TNRKit.jl \cite{TNRKit, TensorKit}.

\paragraph{Contribution} A.U. and S.D.M. contributed equally to this work. A.U. developed the concept of global TNR and its application to thermal TNR, while S.D.M. devised an efficient method for constructing the elementary PEPO via a cluster expansion, which is crucial for improved accuracy.

\bibliography{apssamp}
\clearpage
\onecolumngrid

\begin{center}
\textbf{\Large Supplementary Material}
\end{center}
\vspace{2em}
\twocolumngrid

\appendix
\section{Estimation of transition temperature}
Here, we present the methodology for extrapolating the transition temperature. In constructing the column tensor, we discretize the imaginary-time evolution. As a result of this discretization, the simulation does not always align precisely with the true critical temperature, and the extracted effective central charge deviates from its theoretical value. This deviation becomes amplified under repeated RG transformations and eventually vanishes as the system flows away from criticality. Therefore, we terminate the RG procedure at the 12th step.

\begin{figure}
    \centering
    \includegraphics[width=86mm]{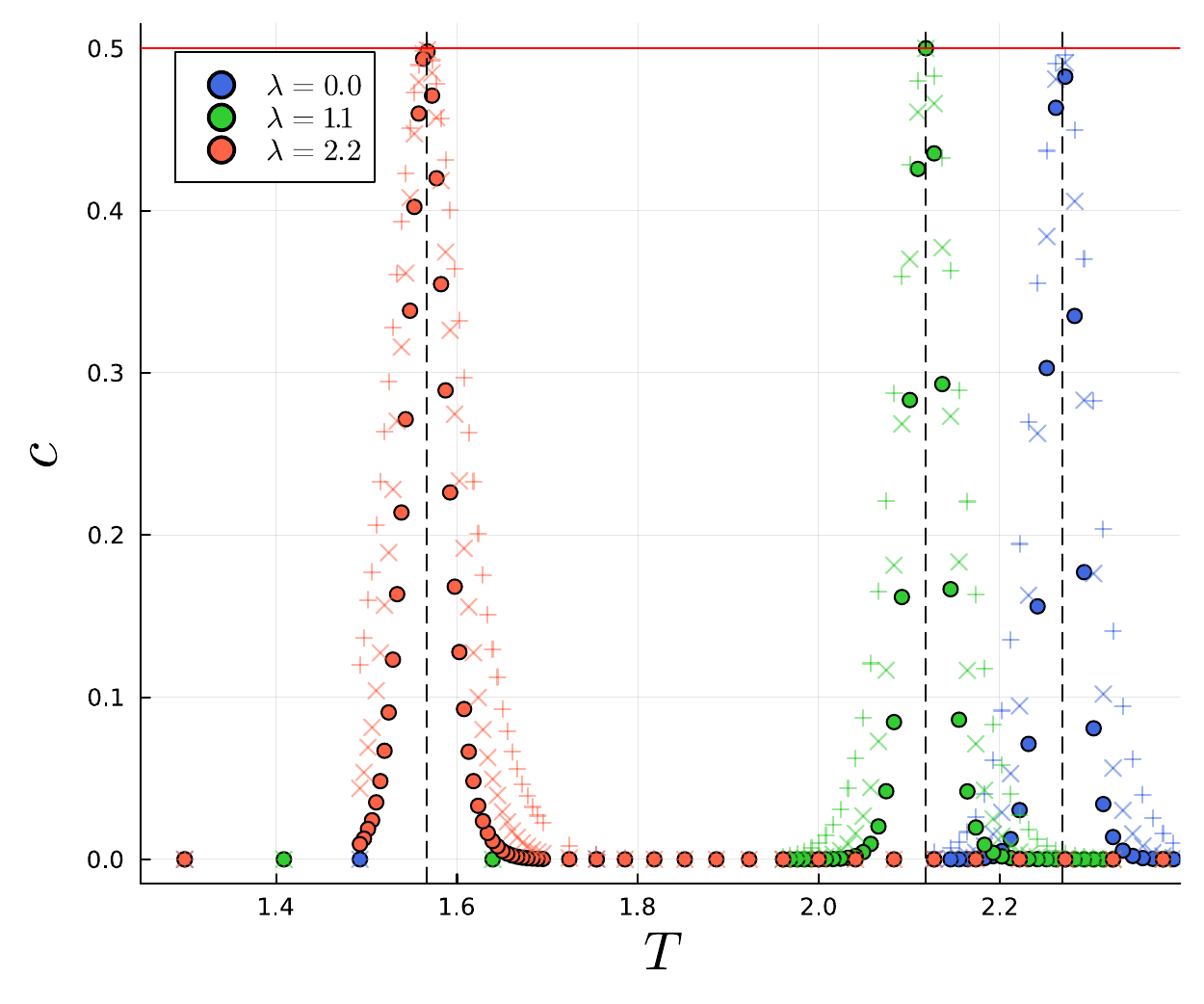}
    \caption{Examples of estimation of transition temperature for $\lambda=0$, 1.1 and 2.2. The cross, x-cross, and solid circles represent, respectively, the effective central charge obtained from 10-12 RG steps. The transition temperatures are determined from the two largest points, as indicated by the black dotted lines.}
    \label{fig:transition}
\end{figure}

Figure~\ref{fig:transition} shows the numerically obtained effective central charge at $\lambda = 0$, 1.1, and 2.2 near their respective critical points. The cross, x-cross, and solid circles denote results after 10, 11, and 12 RG steps, respectively. As the system is coarse-grained, the peak in the effective central charge becomes sharper. Notably, for $\lambda = 0$, the data points lie slightly off the peak and thus exhibit a decay in the central charge with further RG steps.

To estimate the transition temperature, we take the midpoint between the two temperatures corresponding to the largest effective central charge values. In practice, this estimate can be further refined by computing a weighted average of temperatures, where each is weighted by the inverse of its deviation from the ideal central charge value.

\end{document}